\documentclass[12pt,preprint,usenatbib]{aastex}
\usepackage{graphicx,multirow,footnote,color}
\DeclareGraphicsExtensions{.pdf,.png,.eps}

\shorttitle{Sizes of Star Clusters in M83}
\shortauthors{Ryon et al.}

\begin{document}

\title{Sizes and Shapes of Young Star Cluster Light Profiles in M83}

\author{J. E. Ryon}
\affil{Department of Astronomy, University of Wisconsin-Madison, 475 N. Charter St., Madison, WI, 53706, USA}
\email{ryon@astro.wisc.edu}

\author{N. Bastian}
\affil{Astrophysics Research Institute, Liverpool John Moores University, 146 Brownlow Hill, Liverpool L3 5RF, UK}

\author{A. Adamo}
\affil{The Oskar Klein Centre, Department of Astronomy, AlbaNova, Stockholm University, SE-106 91 Stockholm, Sweden}

\author{J. S. Gallagher III}
\affil{Department of Astronomy, University of Wisconsin-Madison, 475 N. Charter St., Madison, WI, 53706, USA}

\author{I. S. Konstantopoulos}
\affil{Australian Astronomical Observatory, PO Box 915, North Ryde NSW 1670, Australia}
\affil{ARC Centre of Excellence for All-Sky Astrophysics (CAASTRO), 44 Rosehill Street, Redfern NSW 2016, Australia}

\author{S. Larsen}
\affil{Department of Astrophysics / IMAPP, Radboud University, P.O. Box 9010, 6500 GL Nijmegen, The Netherlands}

\author{K. Hollyhead}
\affil{Astrophysics Research Institute, Liverpool John Moores University, 146 Brownlow Hill, Liverpool L3 5RF, UK}

\author{E. Silva-Villa}
\affil{Instituto de F\'{i}sica-FCEN, Universidad de Antioquia, Calle 70 No. 52-21, Medellin, Colombia}

\author{L. J. Smith}
\affil{Space Telescope Science Institute and European Space Agency, 3700 San Martin Drive, Baltimore, MD 21218, USA}

\begin{abstract}

We measure the radii and two-dimensional light profiles of a large sample of young, massive star clusters in M83 using archival HST/WFC3 imaging of seven adjacent fields. We use GALFIT to fit the two-dimensional light profiles of the clusters, from which we find effective (half-light) radii, core radii, and slopes of the power-law (EFF) profile ($\eta$). We find lognormal distributions of effective radius and core radius, with medians of $\approx$2.5~pc and $\approx$1.3~pc, respectively. Our results provide strong evidence for a characteristic size of young, massive clusters. The average effective radius and core radius increase somewhat with cluster age. Little to no change in effective radius is observed with increasing galactocentric distance, except perhaps for clusters younger than 100~Myr. We find a shallow correlation between effective radius and mass for the full cluster sample, but a stronger correlation is present for clusters 200-300 Myr in age. Finally, the majority of the clusters are best fit by an EFF model with index $\eta\lesssim3.0$. There is no strong evidence for change in $\eta$ with cluster age, mass, or galactocentric distance. Our results suggest that clusters emerge from early evolution with similar radii and are not strongly affected by the tidal field of M83. Mass loss due to stellar evolution and/or GMC interactions appear to dominate cluster expansion in the age range we study.

\end{abstract}


\section{Introduction}
\label{intro}

Star clusters are an important end result of the star formation process. Because they are gravitationally-bound objects, they can potentially survive for a long time and act as tracers of past star formation events. Due to their high densities and surface brightnesses, massive star clusters can also be detected out to large distances. In fact, with the \textit{Hubble Space Telescope} (HST), individual stars in clusters can be resolved beyond the Local Group  \citep[$\sim$5~Mpc away,][]{larsen2011}, and the extended light profiles of clusters can be resolved out to a few tens of Mpc \citep[e.g.,][]{whitmore1999, schweizer2004, bastian2013}. Star clusters can also be approximated as simple stellar populations, so their properties (e.g., age and mass) can be determined fairly well from integrated photometry \citep[e.g.,][]{cabreraziri2014}. A significant fraction of stars are known to form in clustered environments \citep{lada2003}, and some fraction of these end up in bound clusters, depending on the intensity of star formation \citep{goddard2010,adamo2011,adamo2015}. Understanding cluster evolution and dissolution, which are connected to their sizes and concentrations, is therefore important for understanding star formation in general.

Populations of young massive star clusters (YMCs) in nearby galaxies have been the focus of many studies in recent years. Particularly, there is much interest in understanding whether these objects are young analogs of globular clusters (GCs). If so, we could study this early mode of star formation at much higher resolution locally, and better understand how GCs relate to their host galaxies. One intriguing connection between YMCs and GCs is the apparent uniformity in size across a wide range of age, mass, and environment. Studies of star clusters in the Milky Way \citep[e.g.,][]{harris1996, portegieszwart2010}, Local Group \citep[e.g.,][]{elson1987, mackey2003a, barmby2007}, and other nearby galaxies \citep[e.g.,][]{larsen2004, jordan2005, scheepmaker2007} have found the typical effective radius, or half-light radius, to be 2-3~pc. The lack of a significant relationship between cluster mass and radius \citep{zepf1999,larsen2004,scheepmaker2007} is especially puzzling, given that giant molecular clouds (GMCs) do have such a relation \citep{larson1981}. In addition, cluster age is apparently not strongly correlated with radius \citep[e.g.,][]{larsen2004}, and the findings are mixed for galactocentric distance versus radius \citep[e.g.,][]{vandenbergh1991,jordan2005}. The core radii of clusters can also be studied out to distances of a few Mpc \citep{schweizer2004}, and the spread of cluster core radii has been found to increase significantly with increasing age \citep{elson1991, mackey2003a, glatt2009}. Finally, the power-law slope of young cluster light profiles have been measured in nearby galaxies, and some evidence of steepening or truncation in the outer profile wings with increasing age has been found \citep{whitmore1999, larsen2004, schweizer2004}.

M83 is a face-on spiral galaxy \citep[$i=27^{\circ}$, from the major to minor isophotal diameter ratio given in][]{devaucouleurs1991} located at 4.5~Mpc \citep{thim2003}. Several recent studies have scrutinized many aspects of M83's substantial young cluster population \citep{chandar2010,bastian2011, bastian2012a, silvavilla2013,silvavilla2014,chandar2014, hollyhead2015, adamo2015}. In particular, \cite{bastian2012a} measured the effective radii of a sample of young clusters located in two adjacent HST fields. The authors performed a careful visual inspection of their clusters to ensure that the majority of their objects are older than a crossing time, which implies that they are bound systems. The distribution of radii is lognormal, peaking at $\sim$2.5~pc, and a slight expansion is seen with age. 

In this paper, we build upon \cite{bastian2012a} by studying the sizes and light profiles YMCs in M83 in greater detail. We carefully select massive, isolated clusters with well-behaved light profiles from seven adjacent HST fields to fit with GALFIT \citep{peng2002,peng2010}. This sample is one of the most extensive from a single galaxy for which the slope of the light profile was left as a free parameter rather than set to a fixed value. We compare the effective radii, core radii, and slopes of the power-law light profiles of our clusters to their ages, masses, and galactocentric distances to probe the physical processes that dominate their structural evolution. We compare the results from our sample to other studies of YMC and GC sizes in the Milky Way, Local Group, and other nearby galaxies.

This paper is organized as follows. We describe our HST observations and star cluster selection criteria in Section~\ref{obs-cat}. In Section~\ref{measuring}, we describe how we have measured the sizes and light profiles of the star clusters in our sample. In Section~\ref{results}, we present the effective radii, core radii, and light profile slopes of the clusters and how they compare to other cluster properties. We discuss the implications of these results in Section~\ref{discussion}, and finally, we summarize this study in Section~\ref{conclusions}.

\section{Observations \& Cluster Catalogue}
\label{obs-cat}

We use archival HST/WFC3 imaging data retrieved from the Mikulski Archive for Space Telescopes (MAST), which were calibrated and drizzled on-the-fly when requested for download. These data were obtained by two programs, 11360 (PI: O'Connell) and 12513 (PI: Blair). In total, seven fields were observed across the disk of M83 between August 2009 and September 2012. See Figure~1 in \cite{silvavilla2014} for the location of the HST fields superimposed on M83. All seven fields were imaged in the F336W, F438W, F547M, and F814W filters, except Field 1, for which F555W replaced F547M.

The cluster catalogue from which our sample is selected is presented in \cite{silvavilla2014}, further discussed in \cite{adamo2015}, and briefly summarized here. First, sources were selected from the F547M band in Fields 2 through 7 and F555W in Field 1 using Source Extractor \citep{bertin1996}. A concentration index (CI) cut in F547M/F555W was applied to remove unresolved sources. The CI is the magnitude difference between aperture radii of 1 and 3 pixels, and it provides a measure of the relative size of an object. The CI of an extended object is larger than that of an individual star, so applying a CI cut is one method for removing stars from a star cluster catalogue \citep[e.g.,][]{holtzman1996, whitmore2010}. Sources located within 500~pc of the center of M83 were excluded due to incompleteness because of crowding and regions of high extinction. Each resolved source was visually inspected and classified as either a bona fide cluster (class 1), group/association (class 2), or false detection (class 3). Aperture photometry in the remaining filters was performed on the catalogue. For each source with photometric detections and moderate errors ($<$0.3~mag) in all four filters, the SED-fitting technique presented in \cite{adamo2010a} was used to determine the age, mass, and extinction. 

In this study, we use the F547M/F555W images for fitting the cluster light profiles. These filters are not as strongly affected by dust as compared to the bluer filters, nor by bright (red) stars as compared to F814W. We limit our cluster sample to those sources in the \cite{silvavilla2014} catalogue of mass $\geq$$10^4$~M$_{\odot}$ and visual inspection class 1 (i.e., the most centrally concentrated and isolated cluster candidates; i.e., bona fide clusters). The mass limit minimizes the effects of stochastic sampling of the stellar IMF, which can strongly affect the ages and masses derived from broad-band magnitudes of low mass clusters, but diminishes at high masses \citep[e.g.,][]{popescu2010,fouesneau2010,silvavilla2011}. Limiting our sample to clusters of class 1 ensures that their light profiles are well-resolved and reasonably well-behaved, which is optimal for measuring radii.

Finally, a number of tests showed that clusters without extraneous sources within the fitting region, such as bright stars or background fluctuations, were the clusters that returned the most reliable, stable fits\footnotemark \footnotetext{Also discussed on the GALFIT website: http://users.obs.carnegiescience.edu/peng/work/galfit/TFAQ.html\#sensitivity}. We therefore perform an additional visual inspection step to determine the degree to which a cluster is isolated within the $30 \times 30$~pixel fitting region. To do this, we displayed circles of radius 15 pixels centered on each cluster that satisfied the mass and class limits discussed previously. An isolation flag, called $f_{\mathrm{iso}}$, was assigned to each cluster based on the number and brightness of sources within the circle. The flags are defined as follows:
\begin{itemize}
\item[1.] Isolated. No nearby sources, or sources are faint enough to be indistinguishable from sky background.
\item[2.] Somewhat isolated. One or a few faint sources, or slight fluctuations in sky background.
\item[3.] One or two bright sources nearby.
\item[4.] Crowded. Surrounded by several bright sources.
\end{itemize}
In the analysis that follows, we report GALFIT results for only those clusters with $f_{\mathrm{iso}}=1$ or 2. Limiting our sample to relatively isolated clusters removes most of the youngest clusters (less than a few tens of Myr). Such clusters tend to still be associated with their natal star-forming regions, and are therefore surrounded by young, bright stars that may be unrelated to the clusters themselves \citep{larsen2004}. The number of clusters remaining in the sample after the mass cut, visual inspection class cut, and isolation flag cut is 509.

Our cluster sample is undoubtedly incomplete. It was selected using complicated methods from images that cover a limited region of M83 and vary in exposure time. The clusters we study range in age, mass, surface brightness, and are subject to different local environments (e.g., crowding, dust lanes), though we have attempted to select only the best candidates for this study. For these reasons, the completeness of our sample is nearly impossible to quantify, but we indicate the ways in which we expect our sample to be incomplete in the following sections.

A small amount of processing was required to prepare the HST images for use with GALFIT. First, GALFIT prefers input images to be in units of counts or electrons instead of flux units. We multiplied the drizzled images by the exposure time in order to convert from electrons per second to electrons. Because GALFIT multiplies by the GAIN keyword value in the image header to convert an input image from counts to electrons, and our images were already in units of electrons, we set the GAIN keyword equal to 1.0~$\mathrm{e^-}$/ADU. We also updated the image headers to include the approximate WFC3/UVIS readnoise, 3.11~$\mathrm{e^-}$, in the header keyword RDNOISE.

\section{Measuring Star Cluster Sizes}
\label{measuring}

\subsection{Methodology}
\label{methodology}

We use the two-dimensional fitting package GALFIT \citep{peng2002, peng2010} to fit the light profiles of the star clusters. Since the clusters are relatively young, we assume that an EFF profile (also known as a Moffat profile) accurately describes their true light profiles \citep{elson1987, larsen1999b, whitmore1999, mackey2003a}. A comparison between fits to EFF, King \citep{king1966}, and Wilson \citep{wilson1975} profiles was performed by \cite{mclaughlin2005} for YMCs and GCs in the Milky Way, Fornax dwarf galaxy, and Magellanic Clouds. They found that the more extended profiles, EFF and Wilson, fit young and old clusters better than the King model, in most cases. For most young clusters, the EFF profile fits about as well as the Wilson profile, though in some cases it does worse. We choose to go with the simpler EFF profile because the Wilson profile is not included in GALFIT, and \cite{mclaughlin2005} also find that most of the physical properties of most clusters are well-constrained no matter the choice of light profile.

The EFF profile takes the form
\begin{equation}
\mu(r) = \mu_{0}(1+r^2/a^2)^{-\eta}
\end{equation}
where $\mu$ is the surface brightness in a given band, $a$ is a characteristic radius, and $\eta$ is the power-law exponent that determines the steepness of the profile wings. Note that $\eta$ here is equal to $\gamma/2$ in Equation~1 of \cite{elson1987}. The core radius of a circular EFF profile, which is the radius at which the surface brightness is half its central value, is given by
\begin{equation} \label{eqrcore}
r_{\mathrm{core}} = \mathrm{FWHM}/2 = 2a\sqrt{2^{1/\eta} - 1}.
\end{equation}
The effective radius ($r_{\mathrm{eff}}$), or half-light radius, is defined to be the radius of the circular area which contains half of the total surface brightness of the light profile. This can be written as
\begin{equation} \label{eqreff}
 r_{\mathrm{eff}} = 2r_{\mathrm{core}}\ \frac{\sqrt{(1/2)^{\frac{1}{1-\eta}} - 1}}{2\sqrt{2^{1/\eta} -1}}. 
\end{equation}
 For an elliptical profile, the true core and effective radius can be found by multiplying Equations~\ref{eqrcore} and \ref{eqreff} by a factor of $0.5(1+b/a)$, where $b/a$ is the semiminor to semimajor axis ratio (\texttt{ishape} manual, \citealt{larsen1999b}).

For each cluster, we fit two components, an EFF component and a sky background component. Both are fit simultaneously over a $30 \times 30$~pixel box ($\approx$$26 \times 26$~pc; 1 WFC3/UVIS pixel $\approx$0.87~pc at the distance of M83). We assume the sky background is flat, but leave the amplitude of the sky as a free parameter. For the EFF component, all of the parameters are left free. These parameters are the x and y image coordinates of the cluster center, total magnitude, FWHM, $\eta$, axis ratio ($b/a$), and position angle. Table~\ref{input} details the initial guesses for each of the free parameters, excluding the x and y image coordinates which come directly from the \cite{silvavilla2014} catalogue. GALFIT returns the best-fit values for each free parameter and their $1\sigma$ uncertainties. GALFIT returned an error for 31 out of the 509 clusters in our sample, so these objects were excluded from the following analysis.

\begin{table}
\begin{center}
\caption{GALFIT Input Parameters \label{input}}
\begin{tabular}{ccc}
Parameter & Value \\
\tableline
Total Magnitude & 20.0 mag \\
FWHM & 2.5 pix \\
$\eta$ & 1.5 \\
Axis Ratio & 1.0 \\
Position Angle & $25^{\circ}$ \\
Sky Background & 300.0 $\mathrm{e^{-}}$ \\
\tableline
\end{tabular}
\end{center}
\end{table}

In order to extract structural components from images, GALFIT convolves a model image with a PSF and compares the result to the data. An accurate PSF is therefore essential for reproducing the effects of the telescope optics in the model images. We create PSFs for each field and filter combination from several bright, isolated stars in each image by using \texttt{pstselect} and \texttt{psf} within DAOPHOT in IRAF. We spatially subsample these empirical PSFs by a factor of 10.

\subsection{Testing the Initial Guesses}
\label{testing}

We investigate the effect of varying the input parameters by performing tests on the most isolated, well-behaved clusters in Field 5 ($f_{\mathrm{iso}}=1$). We varied one of the following initial guesses at a time: FWHM, $\eta$, and size of the fitting box. The remaining input parameters were kept the same as in Table~\ref{input}. For each cluster, we calculated the difference between the best-fit FWHM and $\eta$ from the GALFIT run that used the standard initial guesses and the runs in which we varied a single parameter. We denote these differences $\Delta$FWHM and $\Delta\eta$. The average and standard deviation of these differences (e.g., $\langle \Delta \mathrm{FWHM}\rangle$ and $\sigma \Delta \mathrm{FWHM}$) are listed in Table~\ref{test} for each varied input parameter. We find that changing the initial guesses and the fitting box size have little affect on the best-fit FWHM. The average FWHM differences are close to zero, and the scatter about the average is small. The average differences of the best-fit $\eta$ are also close to zero for all initial guesses an box sizes, but the scatter tends to be larger than that of the best-fit FWHM. Changing the fitting box size also results in average differences close to zero, but there is a larger scatter about those averages, especially for $\eta$. Therefore, although changing the fitting box size or initial guesses may affect fits to individual clusters, on the whole, the best-fit results are essentially the same.

\begin{table}[h]
\begin{center}
\caption{Tests of GALFIT Input Parameters \label{test}}
\begin{tabular}{c | c | cccc}
\tableline
Input & $N_{\mathrm{clusters}}$ & \multicolumn{4}{c}{Output} \\
FWHM (pix) & & $\langle\Delta\mathrm{FWHM}\rangle$ (pix) & $\sigma\Delta\mathrm{FWHM}$ (pix) & $\langle\Delta\eta\rangle$ & $\sigma\Delta\eta$ \\
\tableline
1.5 & 86 & -0.007 & 0.08 & 0.01 & 0.4 \\ 
3.5 & 86 & -0.01 & 0.09 & 0.006 & 0.4 \\
\tableline
$\eta$ & & & & & \\
\tableline
1.1 & 86 & 0.03 & 0.1 & 0.1 & 0.5 \\
2.0 & 86 & -0.005 & 0.08 & -0.04 & 0.2 \\
\tableline
Fitting Box (pix) & & & & & \\
\tableline
20$\times$20 & 84 & 0.04 & 0.3 & 0.05 & 0.8 \\
40$\times$40 & 84 & -0.01 & 0.2 & 0.003 & 0.6 \\
\tableline
\end{tabular}
\end{center}
\end{table}

\section{Results}
\label{results}

In this section, we compare the structural parameters from GALFIT with other cluster properties. We also present these values in Table~\ref{results_table} for the 478 clusters that were well-fit by GALFIT. As discussed in Sections~\ref{reff} and \ref{rcore}, the effective radii of clusters best-described by $\eta\leq1.3$ are not well-constrained, and all structural parameters for clusters best-fit by $\eta\leq1.1$ should be treated with caution.

\begin{deluxetable}{ccccccccccccccccccc}
\setlength{\tabcolsep}{0.015in} 
\tabletypesize{\scriptsize}
\rotate
\tablewidth{0pt}
\tablecaption{Properties of YMCs in M83\label{results_table}}
\tablehead{
\colhead{Cluster}  & \colhead{R.A.} & \colhead{DEC} & \colhead{$d_{\mathrm{gc}}$} & \colhead{FWHM} & \colhead{FWHM} & \colhead{$\eta$} & \colhead{$\eta$} & \colhead{$\log r_{\mathrm{eff}}$} &\colhead{$\log r_{\mathrm{eff}}$} & \colhead{$\log r_{\mathrm{core}}$} & \colhead{$\log r_{\mathrm{core}}$} & \colhead{$\log t_{\mathrm{age}}$} & \colhead{Min} & \colhead{Max} & \colhead{$\log M$} & \colhead{Min} & \colhead{Max} & \colhead{$f_{\mathrm{iso}}$} \\
\colhead{ID} & \colhead{(deg)} & \colhead{(deg)} & \colhead{(kpc)} & \colhead{(pix)} & \colhead{Error (pix)} & & \colhead{Error} & \colhead{(pc)} & \colhead{Error (pc)} & \colhead{(pc)} & \colhead{Error (pc)} & \colhead{(yr)} & \colhead{$\log t_{\mathrm{age}}$ (yr)} & \colhead{$\log t_{\mathrm{age}}$ (yr)} & \colhead{(M$_{\odot}$)} & \colhead{$\log M$ (M$_{\odot}$)} & \colhead{$\log M$ (M$_{\odot}$)} &
} 
\startdata
10016 & 204.26505 & -29.88305 & 1.5973 &  5.22 & 0.28 &  1.25 &  0.07 &  0.99 &   0.15 &  0.33 &  0.04 &  9.10 &  8.20 &  9.10 & 4.56 & 4.37 & 4.60 & 1 \\
10029 & 204.27220 & -29.88226 & 1.8411 &  2.09 & 0.10 &  1.14 &  0.03 &  1.06 &   0.21 & -0.06 &  0.03 &  9.30 &  8.10 &  9.70 & 4.92 & 4.64 & 5.14 & 1 \\
10083 & 204.28027 & -29.87885 & 2.1126 &  3.97 & 0.36 &  1.66 &  0.16 &  0.48 &   0.06 &  0.20 &  0.05 &  9.00 &  8.96 &  9.10 & 4.17 & 4.12 & 4.24 & 1 \\
10086 & 204.25021 & -29.87869 & 1.0754 &  4.97 & 0.54 &  2.49 &  0.37 &  0.46 &   0.06 &  0.33 &  0.05 &  7.90 &  7.70 &  8.00 & 4.11 & 4.01 & 4.17 & 1 \\
10090 & 204.28065 & -29.87853 & 2.1231 &  3.89 & 0.50 &  1.01 &  0.10 & 15.28 & 149.35 &  0.22 &  0.06 &  8.78 &  8.42 &  8.90 & 4.28 & 4.21 & 4.34 & 1 \\
\enddata
\tablecomments{Col. (1): Cluster ID number. Cols. (2) and (3): R.A. and Dec coordinates in decimal degrees. Col. (4): Galactocentric distance in kiloparsecs. Cols. (5) and (6): FWHM of the EFF light profile in pixels and the 1$\sigma$ error as reported by GALFIT. Cols. (7) and (8): Power-law index, $\eta$, of the EFF light profile and the 1$\sigma$ error as reported by GALFIT. Cols. (9) and (10): Log of the half-light (effective) radius in parsecs and the 1$\sigma$ error. Cols. (11) and (12): Log of the core radius in parsecs and the 1$\sigma$ error. Cols. (13)-(15): Log of the best-fit cluster age in years and associated minimum and maximum ages allowed by the SED fits. Clusters of all ages are presented in this table, but the main results of the paper are based on those $\leq$300~Myr in age. Cols. (16)-(18): Log of the best-fit cluster mass in solar masses and associated minimum and maximum masses allowed by the SED fits. Col. (19): Isolation flag, $f_{\mathrm{iso}}$. \\
As discussed in Sections~\ref{reff} and \ref{rcore}, the effective radii of clusters best-described by $\eta\leq1.3$ are not well-constrained, and all structural parameters for clusters best-fit by $\eta\leq1.1$ should be treated with caution. Note that some of the structural parameters in this table may be different than those in the full catalogue presented in  \cite{silvavilla2015}. \\
Only a portion of this table is shown here to demonstrate its content. A machine-readable version of the full table is available online.}
\end{deluxetable}

\subsection{Effective Radius}
\label{reff}

\begin{figure}[h]
\centering
\includegraphics[scale=0.6]{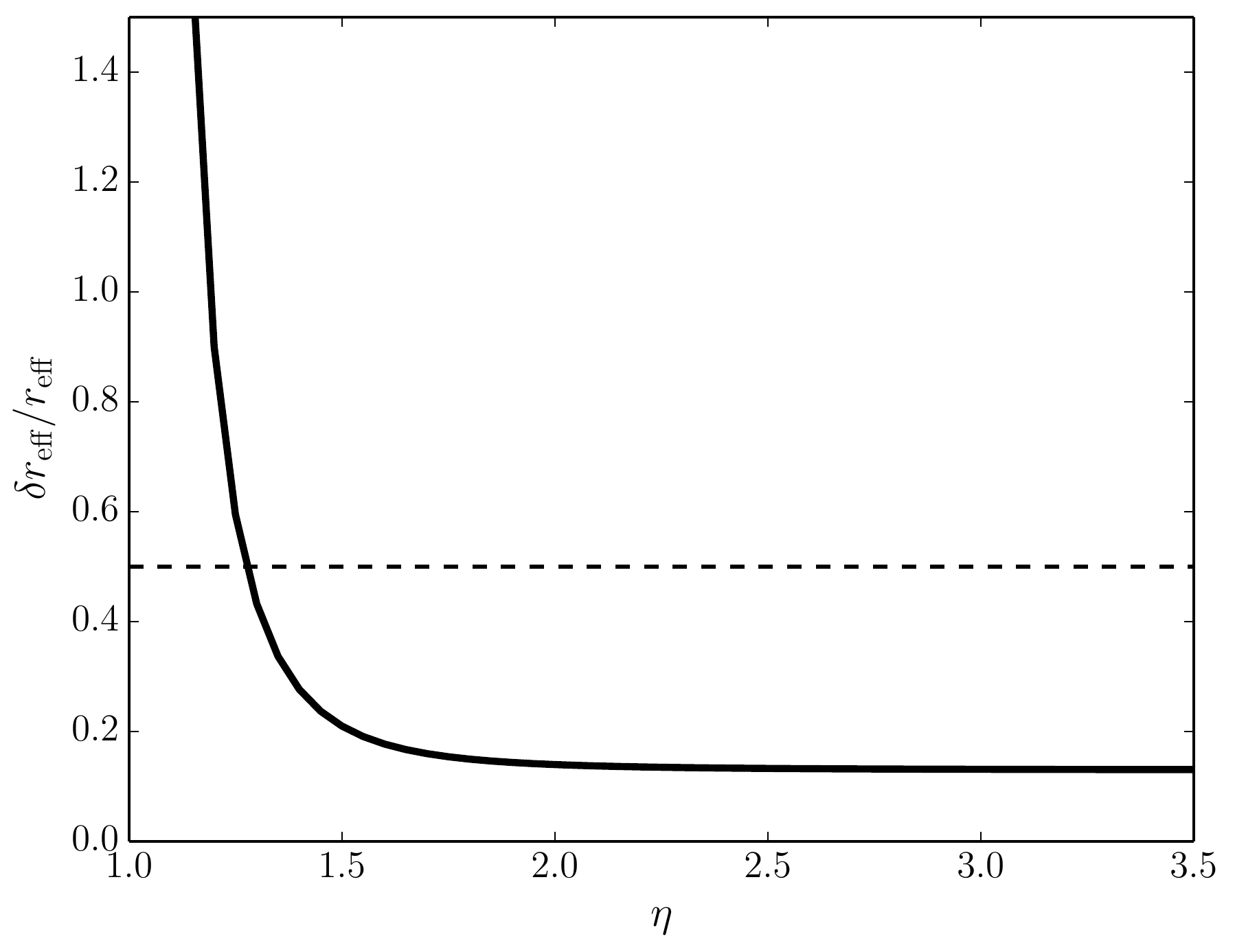}
\caption{Fractional error on the effective radius as a function of the power-law index, $\eta$. To calculate $\delta r_{\mathrm{eff}}$, we assume 10\% errors on FWHM, $\eta$, and the axis ratio $b/a$. As shown by the dashed line, the fractional error is $\approx$50\% at $\eta\approx1.3$. \label{fracerr}}
\end{figure}

Measuring $r_{\mathrm{eff}}$ is difficult when the power-law index of the EFF profile, $\eta$, is near 1.0. When $\eta \leq 1.0$, $r_{\mathrm{eff}}$ is undefined (unless an outer truncation radius is assumed, as in \cite{larsen2001} and \cite{larsen2004}). When $\eta$ is slightly greater than 1.0, $r_{\mathrm{eff}}$ diverges to unphysical values, as does the error, $\delta r_{\mathrm{eff}}$, as shown in Figure~\ref{fracerr}. In this plot, we show the fractional error of $r_{\mathrm{eff}}$ with respect to $\eta$, assuming 10\% errors in FWHM, $\eta$, and axis ratio $b/a$. In order to avoid unphysical effective radii and errors, we study clusters that are best fit by $\eta \geq 1.3$, which corresponds to a fractional error in effective radius of $\approx$50\%, in this section. We briefly discuss the biases this $\eta$ limit imposes on our sample in Section~\ref{dist_reff}. The number of clusters remaining after including this $\eta$ limit is 247, which is about half of the 478 clusters that were well-fit by GALFIT.

In addition, we also impose an age limit on our sample of $\leq$300~Myr because the cluster age estimates from SED-fitting become fairly uncertain above this age. Stellar evolutionary models that use different AGB star prescriptions show significant discrepancies around 300~Myr. Therefore, the cluster age determination of the SED fits becomes strongly model dependent at this point. In addition, the age-metallicity degeneracy of globular clusters (GCs) comes into play. Metal-poor GCs a few Gyr old can populate an area in color space that is also occupied by 300~Myr solar-metallicity clusters. The number of clusters with ages $\leq$300~Myr is 152, and the number with ages $>$300~Myr is 95. 

\subsubsection{Distribution of Effective Radii}
\label{dist_reff}

\begin{figure}[t]
\centering
\includegraphics[scale=0.6]{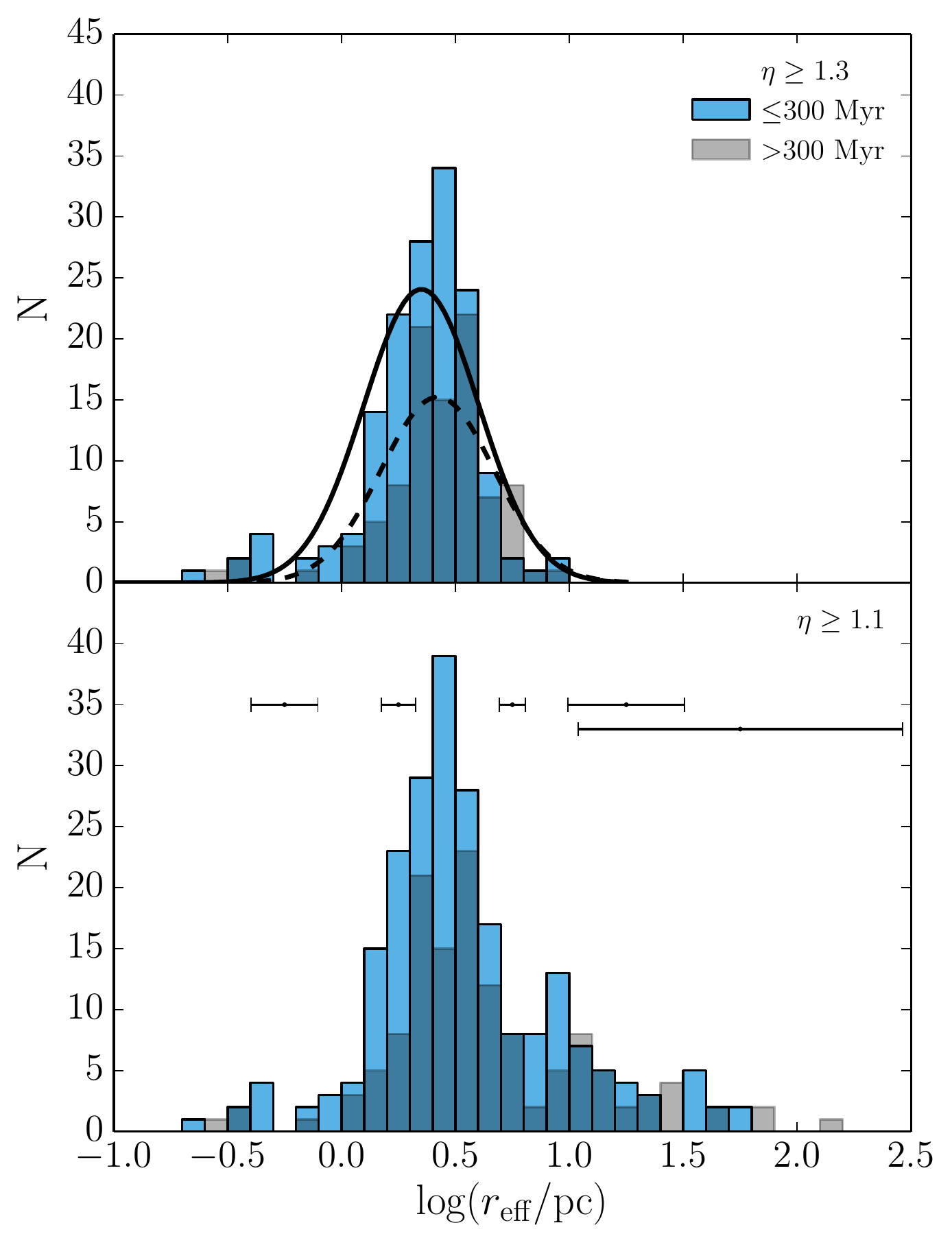}
\caption{Distribution of effective radii. Clusters younger than 300~Myr are plotted in blue and clusters older than 300~Myr are plotted in gray. Top panel: Only clusters best-fit by $\eta\geq1.3$ are included. Lognormal fits to the two age groups are plotted as solid (young clusters) and dashed (old clusters) black lines. Bottom panel: Clusters best-fit by $\eta\geq1.1$. The error bars are median errors of clusters located in 0.5 dex bins in effective radius. \label{reff-hist}}
\end{figure}

In Figure~\ref{reff-hist}, we present the distribution of effective radii for the conservative $\eta$ limit discussed above, $\eta\geq1.3$ (top panel), and a less conservative limit, $\eta\geq1.1$ (bottom panel). Clusters younger than 300~Myr are plotted in blue and clusters older than 300~Myr are plotted in gray. In the top panel, both age groups are well-described by lognormal distributions, as shown by fits to the data plotted as solid and dashed black lines. The median effective radius of the young sample in the top panel is $\approx$2.5~pc, while that for the old sample is $\approx$2.8~pc. The full range of effective radii covered by either sample is $\approx$0.3 to 10~pc.

Including clusters best-fit by smaller $\eta$ values introduces a tail of more extended objects into the distribution, as seen in the bottom panel. The presence (and location) of the peak in the distribution does not change when these extended objects are introduced. In addition, only a few very compact clusters ($\lesssim$0.4~pc) appear in either panel due to the CI cut described in Section~\ref{obs-cat}. We can easily distinguish clusters 1 to 1.5~pc in size from unresolved objects, so the dropoff from the peak to smaller sizes in either panel is not due to biases in the sample. The median effective radius is $\approx$3.0~pc for the young sample and $\approx$3.5~pc for the old sample in the bottom panel. We therefore find strong evidence for a characteristic size of YMCs in M83 of about 2.5 to 3~pc.

We also plot median error bars for each 0.5 dex in effective radius in the bottom panel to show that the extended clusters have much larger errors than the smaller clusters. Therefore, while there does exist a population of extended clusters in M83, the errors are too large to include them in our analysis. In the remainder of Section~\ref{reff}, we only consider clusters described by $\eta\geq1.3$.

The distribution in the top panel of Figure~\ref{reff-hist} is very similar to that found by \cite{bastian2012a} in Fields 1 and 2 of M83 (their Figure~13). This includes the overall lognormal shape, peak radius, and range of radii present. These similarities are clear despite the fact that \cite{bastian2012a} used ISHAPE \citep{larsen1999b} instead of GALFIT, and assumed a fixed $\eta$ of 1.5. It seems that allowing $\eta$ to remain a free parameter does not drastically effect the shape of the distribution. In other words, the effective radius appears to be fairly accurately recovered from a well-behaved light profile by a two-dimensional fit, even if the assumed shape of the light profile is somewhat inaccurate, as previously suggested by \cite{kundu1998}, \cite{larsen1999b}, and \cite{mclaughlin2005}. 

The distribution in Figure~\ref{reff-hist} is also quite similar to that found by studies of young massive clusters \citep{meurer1995, larsen1999b, whitmore1999, larsen2004, bastian2005, lee2005, barmby2006, scheepmaker2007, mayya2008, portegieszwart2010, annibali2011, johnson2012, sanroman2012} and globular clusters \citep{jordan2005,spitler2006,georgiev2008,harris2009,harris2010,masters2010} in the Milky Way and other nearby galaxies.  Some galaxies also harbor populations of extended clusters, such as the ``faint fuzzies'' in NGC~1023 \citep{larsen2000}, the extended globular clusters in M31 \citep{huxor2005}, and the extended, faint clusters in M101 \citep{simanton2015}. The effective radii of these objects range from $\sim$7 to $\sim$30~pc, but they may not be all that different structurally than extended GCs in the Milky Way's outer halo \citep{vandenbergh2004}. The ``faint fuzzies'', however, appear to be a unique population of objects, which have very different kinematics, metallicities, and luminosities than extended Milky Way GCs \citep{brodie2002}. Massive clusters of all ages in the Magellanic Clouds also appear to be somewhat extended compared to those in most other systems, having median effective radii of $\sim$8~pc \citep{mclaughlin2005, glatt2009}. We discuss these populations of extended clusters further in Section~\ref{sim}.

\subsubsection{Effective Radius versus Age}
\label{size-age}

\begin{figure}[t]
\centering
\includegraphics[scale=0.6]{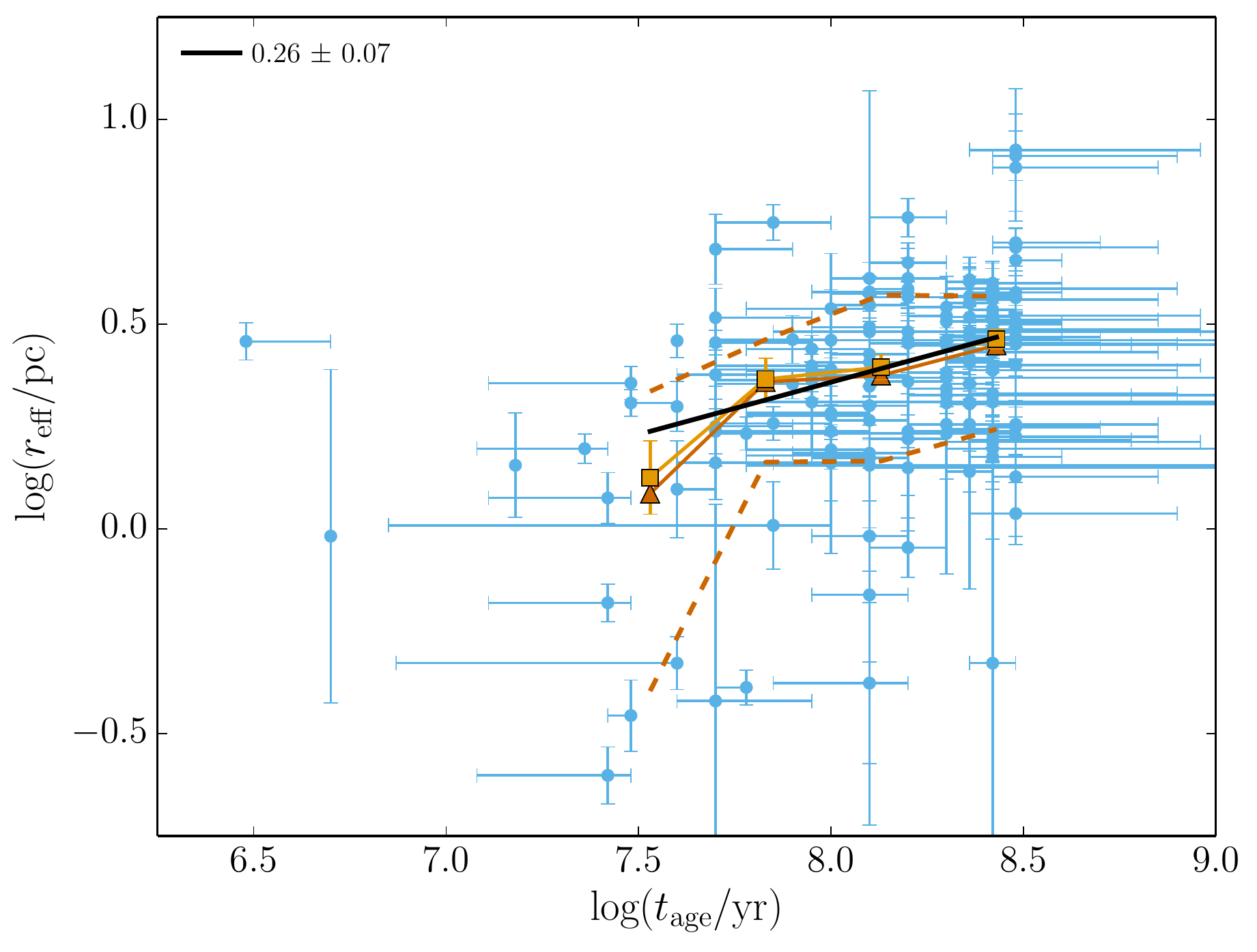}
\caption{Effective radius as a function of cluster age. The blue points are clusters $\leq$300~Myr in age and best-fit by $\eta\geq1.3$. Median and mean effective radii are calculated in 0.3 dex age bins containing 5 or more clusters, and are plotted as orange triangles and yellow squares, respectively. One standard deviation in effective radius about the mean is plotted as dashed orange curves. The black line is a least-squares fit to the mean effective radii, which includes the standard errors on the mean. \label{reff-age}}
\end{figure}

In Figure~\ref{reff-age}, we plot effective radius as a function of age. For bins spanning 0.3 dex in age containing more than 5 clusters, we calculate the median and mean effective radii, which are plotted as orange triangles and yellow squares, respectively. We also plot one standard deviation about the mean for each bin as dashed orange curves. Finally, we perform a least-squares fit to the mean effective radii, including the standard errors on the mean, and plot the result as a black line. We find a clear correlation between mean radius and age with a slope of $0.26\pm0.07$. The mean mass of the clusters in each bin is relatively constant at $\approx$20,000~M$_{\odot}$, so this correlation is unrelated to cluster mass.

As we discuss in Section~\ref{obs-cat}, our cluster sample suffers from incompleteness which is difficult to quantify. The most obvious bias in our sample is the lack of young clusters, $\lesssim$30-50~Myr in age, due to the isolation flag cut discussed in Section~\ref{obs-cat}. Of the young clusters that exist in our sample, we would expect that all sizes are well-represented because young clusters are relatively easy to detect due to their brightness. On the other hand, it is likely that extended, old clusters are not fully represented. The low surface brightness of these objects make them difficult to detect against the background. Thus, the trend in Figure~\ref{reff-age} could actually be steeper if there is a significant population of faint, extended clusters that has gone undetected.

This trend is similar to the size-age correlation found by \cite{bastian2012a}, though our data show larger scatter and cover a smaller range in age. In fact, \cite{bastian2012a} note that the correlation they find may be due to the fixed power-law index of the light profile, but that hypothesis is now ruled out with this study. Several other studies have also looked for a relation between effective radius and age in young massive clusters. \cite{larsen2004} finds essentially no trend for a large sample of clusters from several nearby galaxies. \cite{barmby2009} also find no trend for 23 clusters in M31, and \cite{sanroman2012} find no trend for 161 clusters in M33. The young clusters studied by \cite{mclaughlin2005} in the LMC and SMC show no relationship between age and effective radius. In M51, three studies find a slight positive relationship with a power-law slope of $\sim$0.1 \citep{lee2005, scheepmaker2007, hwang2010}. On the other hand, another study finds a very slight negative trend in M51, though clusters smaller than 2~pc were unresolved and not included in the fit \citep{bastian2005}.

\subsubsection{Effective Radius versus Galactocentric Distance}
\label{size-distance}

\begin{figure}[t]
\centering
\includegraphics[scale=0.6]{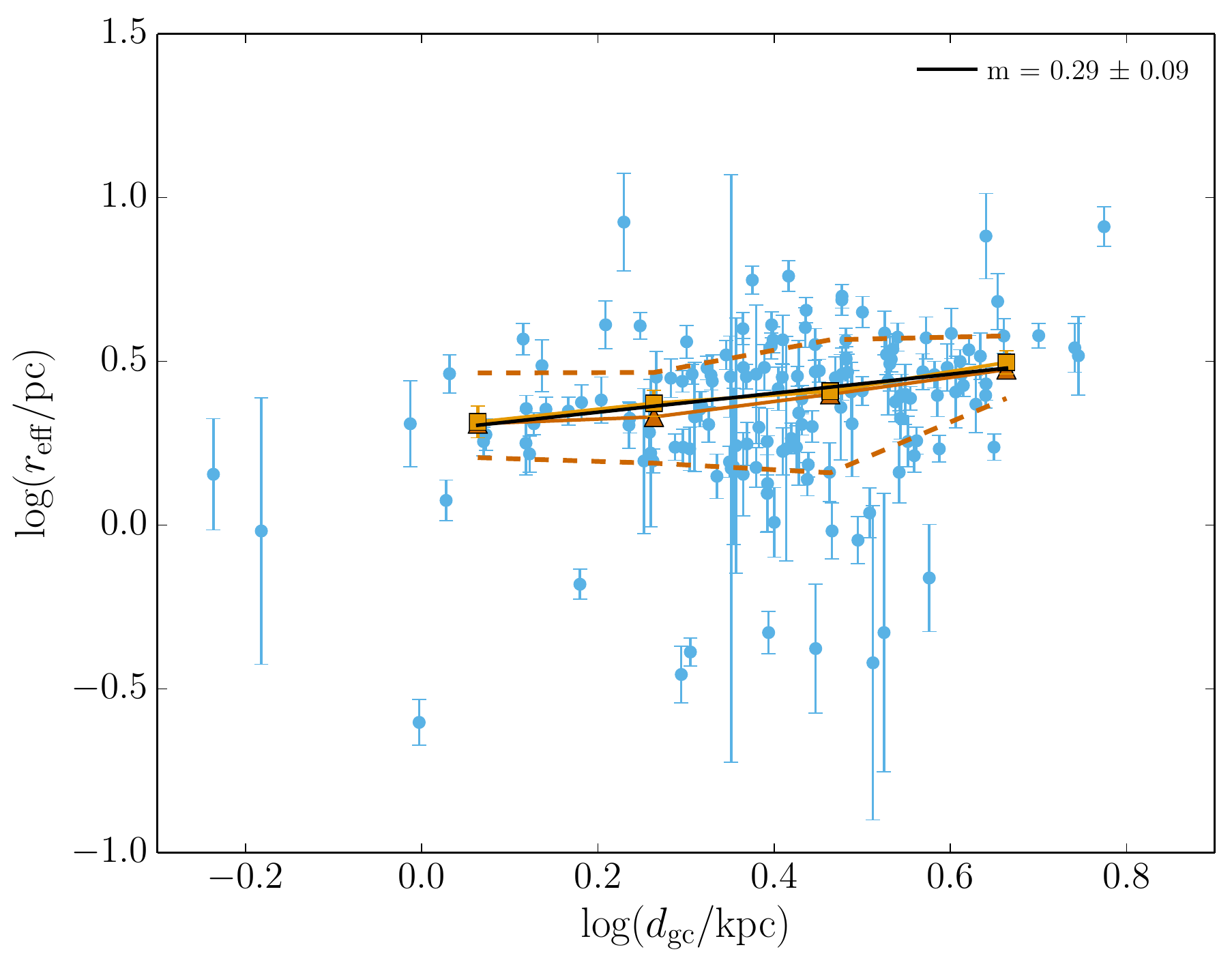}
\caption{Effective radius as a function of galactocentric distance. Points and lines are the same as in Figure~\ref{reff-age}, except that the mean, median, and standard deviation are found for 0.2 dex bins in galactocentric distance. \label{reff-dgc}}
\end{figure}

\begin{figure}[t]
\centering
\includegraphics[scale=0.6]{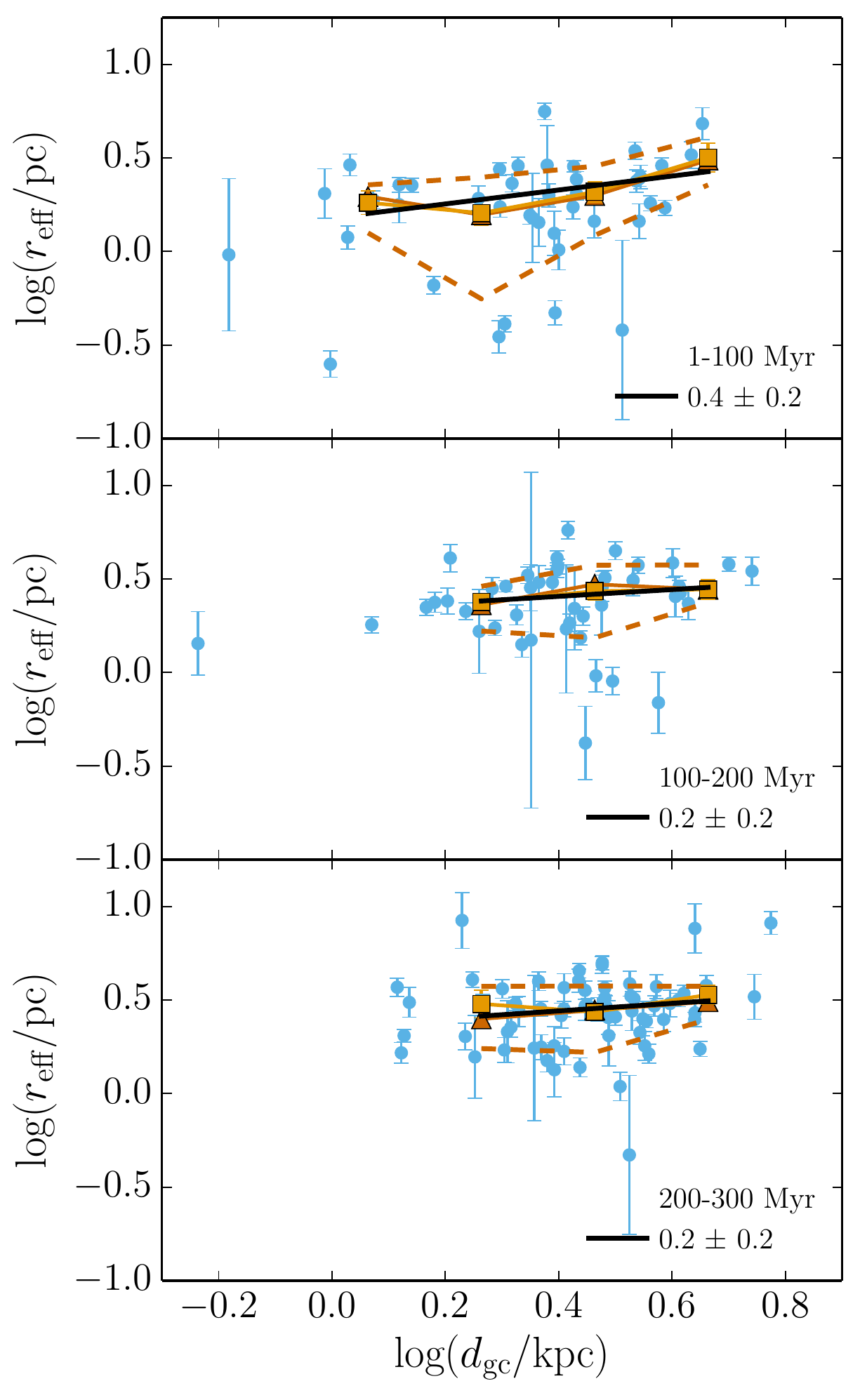}
\caption{Same as Figure~\ref{reff-dgc}, except that the cluster sample is now divided into three bins in age. \label{reff-dgc-agebins}}
\end{figure}

In Figure~\ref{reff-dgc}, we plot effective radius as a function of galactocentric distance. Again, we only include clusters $\leq$300~Myr in age and best-fit by $\eta\geq1.3$. The least-squares fit to the mean effective radius in each 0.2 dex bin finds a correlation with slope 0.29 $\pm$ 0.09. However, cluster age is also weakly correlated with galactocentric distance, such that within 2~kpc of the center of M83, clusters are $\approx$140~Myr old on average. The average age increases to $\approx$230~Myr beyond 4~kpc from the center. We therefore separate the clusters into three 100~Myr bins in age in Figure~\ref{reff-dgc-agebins} to remove the age-distance relation. We perform least-squares fits to the mean effective radii in each age bin, and plot the results as black lines with the slopes labeled in the lower right hand corner of each panel. The bottom two panels show little to no correlation between mean radius and galactocentric distance, while the top panel, containing the youngest clusters, shows a moderate correlation. However, there is considerable scatter in this plot, so it is unclear if the trend is real.

Past studies of young massive clusters have found mixed results with this relationship. \cite{bastian2012a} notes that clusters in Field 2 (further from the center) are on average 0.4~pc larger than those in Field 1 (closer to the center). They attribute this difference to the relationship between cluster age and radius, which corresponds with our results. In M51, the evidence is mixed: \cite{bastian2005} find no obvious trend between projected galactocentric distance and effective radius, but \cite{scheepmaker2007} find a weak relation with a power-law slope of $0.12\pm0.02$. In M33, there is no correlation between galactocentric radius normalized by circular velocity and cluster size, unless a small number of large, outer clusters are included \citep{sanroman2012}. There is a somewhat stronger trend in NGC~7252, where \cite{bastian2013} find a slope of $0.35\pm0.20$ for very massive clusters, $>$$10^6$~M$_{\odot}$. 

For GCs, the picture is almost the same, though most studies cover a much larger range in galactocentric distance than our sample. In the Milky Way, there is a strong relationship between galactocentric distance and radius. \cite{vandenbergh1991} find a power-law stope of $0.5$. In contrast, M31 shows little to no correlation \citep{barmby2007}. Studies of GCs in nearby early-type galaxies find much shallower relations ($\approx$0.1), or almost none at all \citep{jordan2005,madrid2009,harris2009,harris2010,masters2010}, although \cite{spitler2006} find a somewhat stronger relation for a GCs around the Sombrero Galaxy. Further, in dwarf irregular galaxies, \cite{georgiev2008} find no clear trend between galactocentric distance and cluster size.

\subsubsection{Effective Radius versus Mass}
\label{size-mass}

\begin{figure}[t]
\centering
\includegraphics[scale=0.6]{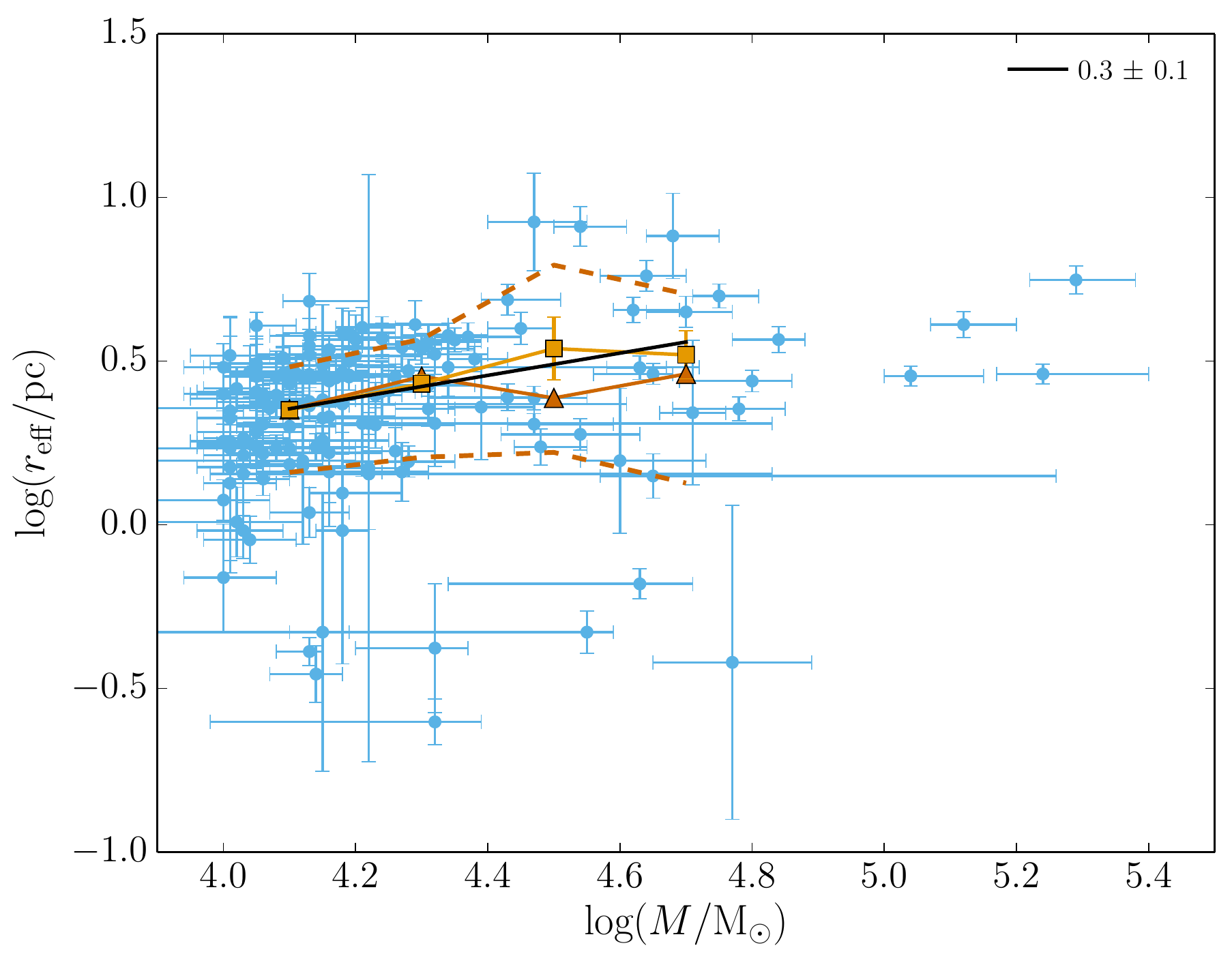}
\caption{Effective radius as a function of cluster mass. Points and lines are the same as in Figure~\ref{reff-age}, except that the mean, median, and standard deviation are found for 0.2 dex bins in mass.  \label{reff-mass}}
\end{figure}

\begin{figure}[t]
\centering
\includegraphics[scale=0.6]{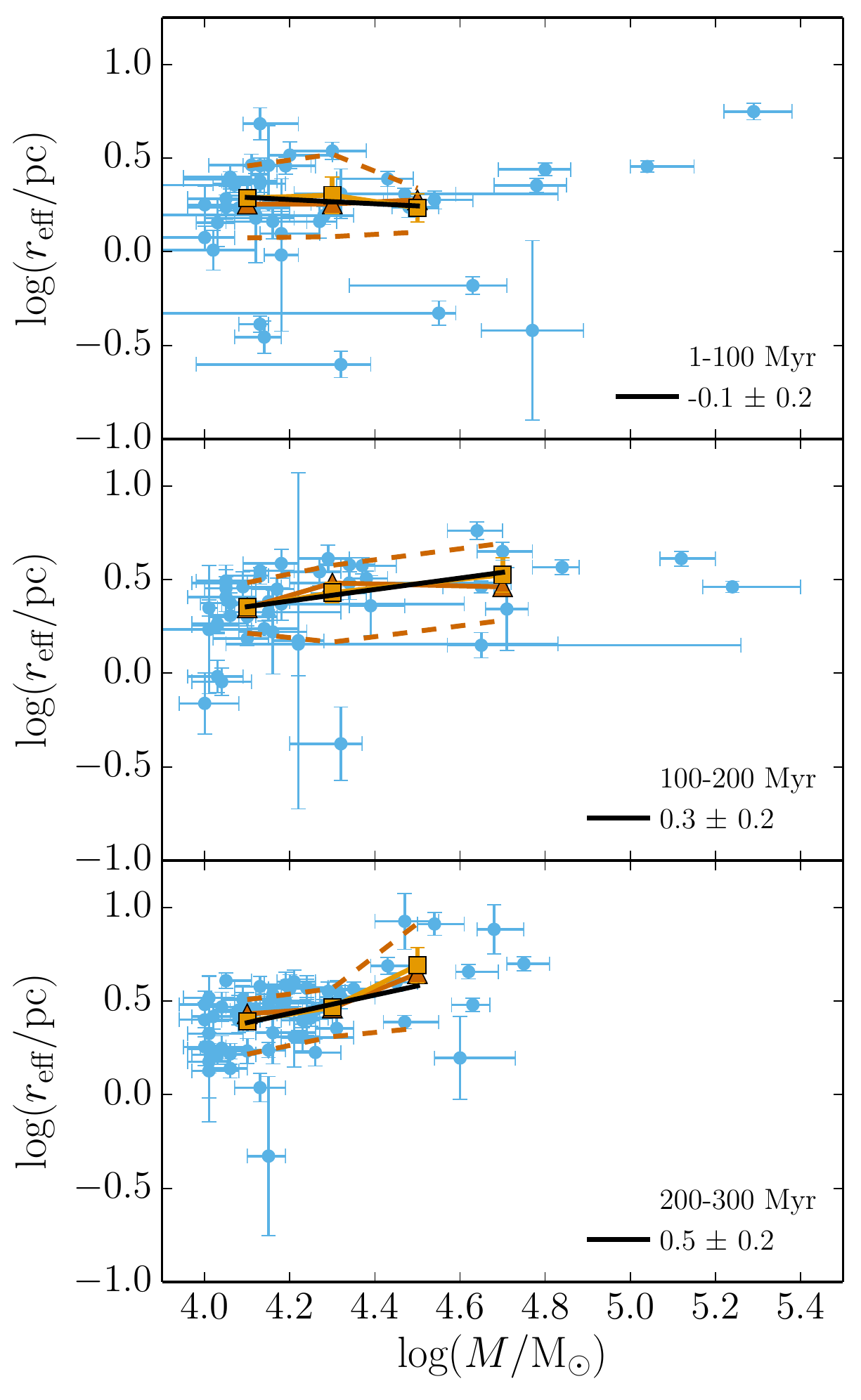}
\caption{Same as Figure~\ref{reff-mass}, except that the cluster sample is now divided into three bins in age. \label{reff-mass-agebin}}
\end{figure}

As discussed in \cite{gieles2006c}, disruption of YMCs by molecular clouds depends on cluster density. Quantifying the mass-radius relation is therefore important for understanding cluster disruption in gas-rich environments like galaxy disks \citep[e.g.,][]{fall2012}. In Figure~\ref{reff-mass}, we plot effective radius as a function of cluster mass. We perform a least-squares fit to the mean effective radii in bins of 0.2 dex in mass and find a shallow correlation with slope 0.3 $\pm$ 0.1. To remove the known age-radius relation shown in Section~\ref{size-age}, we divide the young clusters into three 100~Myr bins in age and perform a least-squares fit to the clusters in each bin, as shown in Figure~\ref{reff-mass-agebin}. The correlation between mass and mean effective radius becomes stronger with increasing cluster age. Notably, the 200-300~Myr bin has a slope of 0.5 $\pm$ 0.2, although the range of cluster masses contained in this bin is fairly narrow and may be skewing the results.

As discussed previously, faint, extended clusters are likely missing from our sample because they are particularly difficult to detect. However, the correlation between mean radius and mass we find for clusters 100-300~Myr in age would not necessarily be diminished unless a large population of low-mass, extended clusters has gone undetected. We expect the sample to be deficient in extended clusters of all masses, not necessarily low-mass clusters alone.

Many studies of young massive clusters have found little to no correlation between effective radius and mass (or luminosity) \citep{meurer1995,zepf1999,larsen2004,lee2005,bastian2005,scheepmaker2007, mayya2008, barmby2009, hwang2010,bastian2012a}. There are a few interesting exceptions, however. The data provided by \cite{mclaughlin2005} for intermediate-age clusters ($>$300~Myr) and GCs in the Magellanic Clouds show a rather strong relation, on the order $r_{\mathrm{eff}} \propto M^{0.5}$ (see Figure~4 in \citealt{fall2012}). \cite{bastian2013} find a power-law slope relation of $0.29\pm0.09$ for the very massive  ($>$$10^6$~M$_{\odot}$) young clusters in NGC~7252, and little to no relation below this mass. \cite{kisslerpatig2006} find a similar relation for a larger sample of very massive young clusters. GCs show little to no relationship between luminosity and radius \citep{vandenbergh1991,kundu2001,madrid2009}, except for perhaps the most luminous (most massive) objects, which may have larger radii \citep{spitler2006, barmby2007,harris2009,harris2010}. Therefore, for all types of clusters, there is essentially no evidence for a strong mass-radius relation below $\sim$$10^6$~M$_{\odot}$, but higher mass objects do appear to get larger with increasing mass.

\subsection{Core Radius}
\label{rcore}

Considering the core radius is advantageous because it is less biased against extended clusters than the effective radius. The best-fit FWHM from GALFIT is equal to twice the core radius, as described in Section~\ref{methodology}. Because the core radius does not depend on $\eta$, unlike the effective radius, we do not have to impose the same $\eta$ limit as we do in Section~\ref{reff}. We relax the limit to $\eta\geq1.1$ in this section for the following reason. GALFIT cannot fit the light profiles of clusters with $\eta <1.0$, which by definition contain an infinite amount of light. Even though this is unphysical, \cite{larsen2004} has shown that clusters with $\eta<1.0$ light profiles do exist. We find that GALFIT assigns these `unphysical' clusters $\eta$ values slightly higher than 1.0. This $\eta$ limit does remove some clusters that are accurately described by $\eta\sim1.0$, but since GALFIT cannot distinguish these clusters from those with unphysically shallow profiles, we remove all clusters best-fit by $\eta<1.1$. With this somewhat relaxed restriction, the number of clusters $\leq$300~Myr in age increases to 224, and the number $>$300~Myr in age is 133.

\begin{figure}[t]
\centering
\includegraphics[scale=0.6]{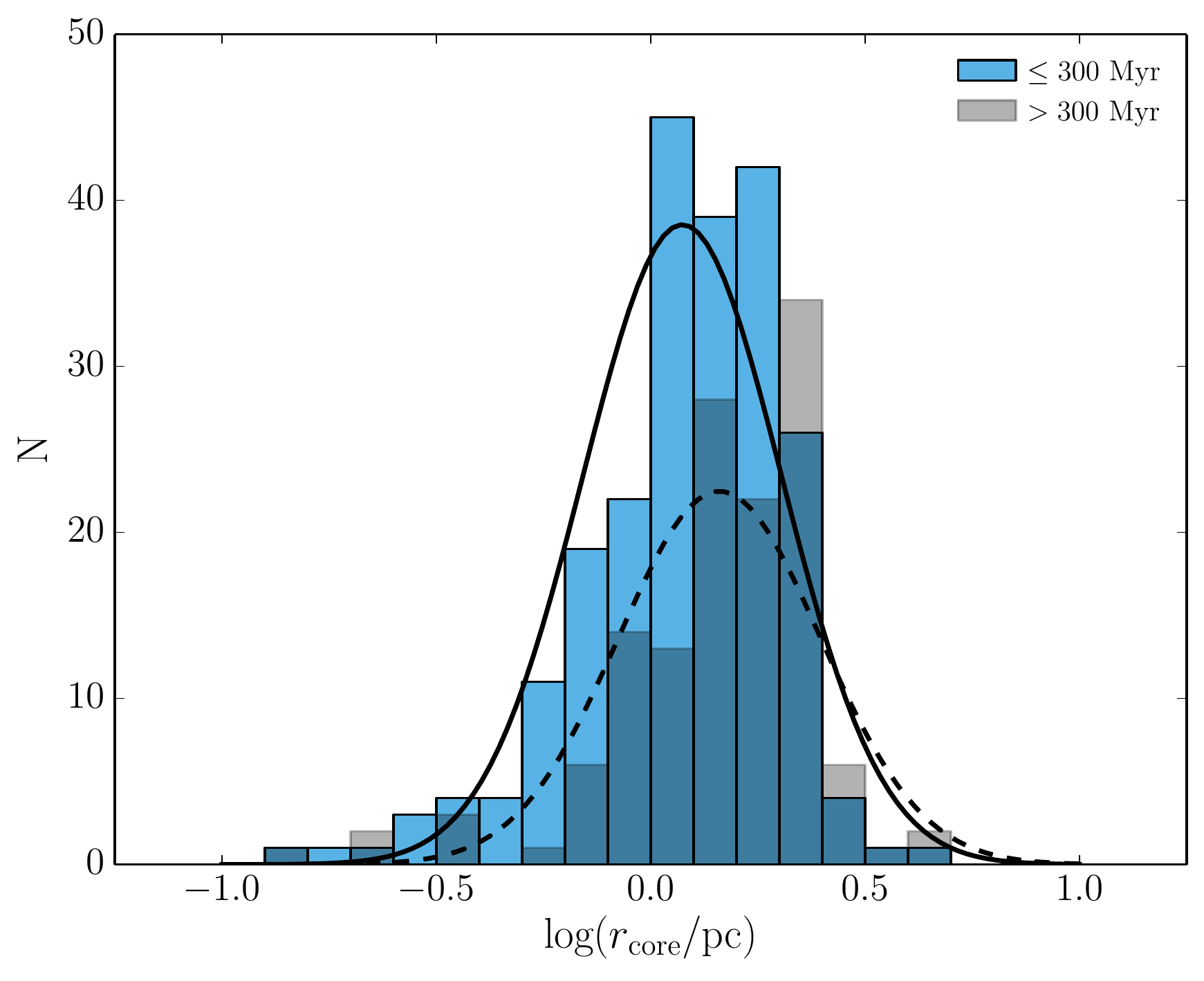}
\caption{Distribution of core radii. Clusters younger than 300~Myr are plotted in blue and clusters older than 300~Myr are plotted in gray. Only clusters best-fit by $\eta\geq1.1$ are included. Lognormal fits to the two age groups are plotted as solid (young clusters) and dashed (old clusters) black lines. \label{rcore-hist}}
\end{figure}

\begin{figure}[t]
\centering
\includegraphics[scale=0.6]{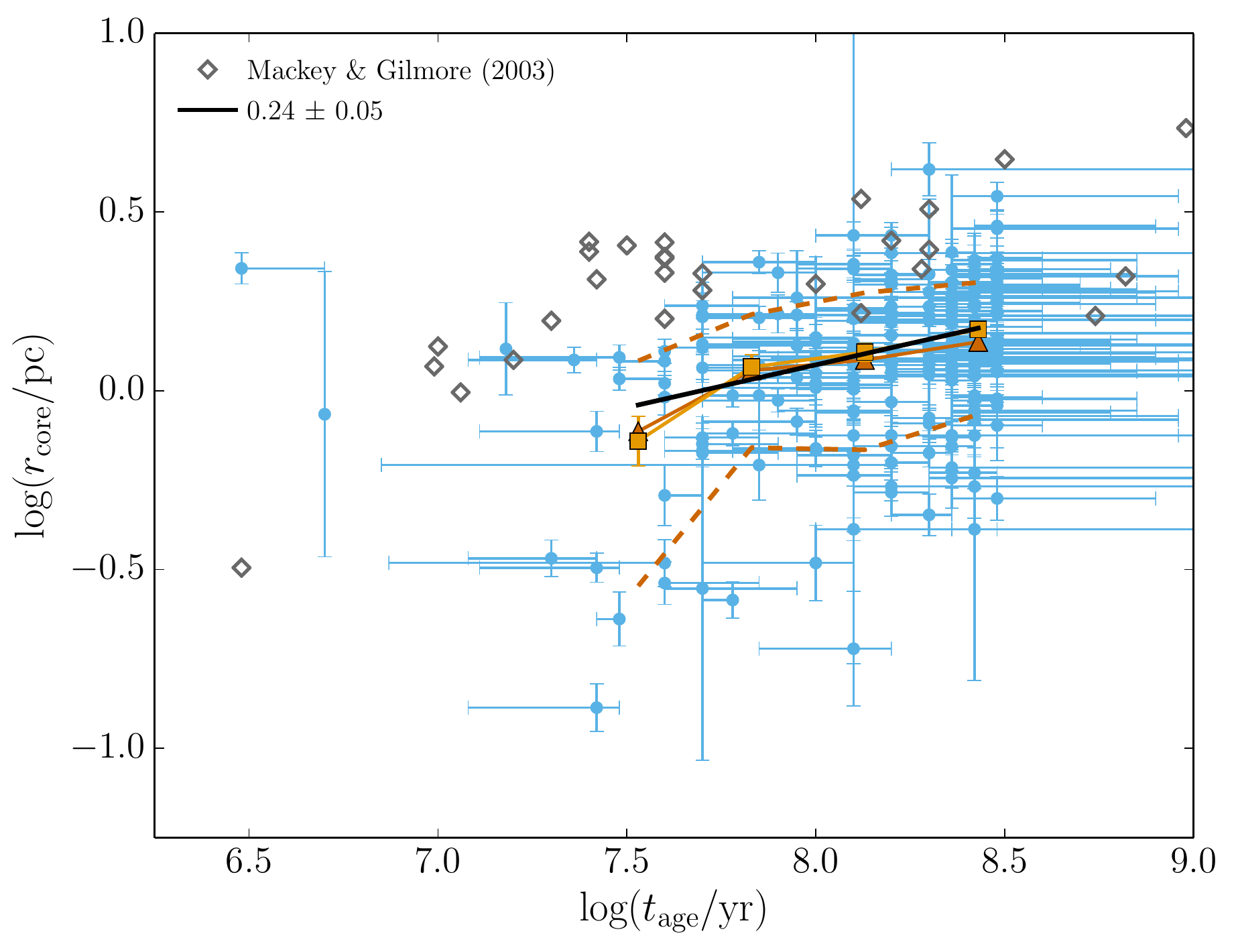}
\caption{Core radius as a function of age. The blue points are M83 clusters $\leq$300~Myr in age and best-fit by $\eta\geq1.1$. The gray diamonds are LMC clusters from \cite{mackey2003a} of similar age to the M83 clusters. Median and mean core radii of the M83 clusters are calculated in 0.3 dex age bins containing 5 or more clusters, and are plotted as orange triangles and yellow squares, respectively. One standard deviation in core radius about the mean is plotted as dashed orange curves. The black line is a least-squares fit to the mean core radii of the M83 clusters, which includes the standard errors on the mean. \label{rcore-age}}
\end{figure}

We plot the distribution of core radii in Figure~\ref{rcore-hist}. Both age groups are fairly well-described by a lognormal distribution (with some skew to the left), as shown by fits to the data represented by solid and dashed black lines. The median core radius of the young sample is $\approx$1.3~pc and that for the old sample is $\approx$1.6~pc. The full range of core radii is $\approx$0.1 to 4~pc. Even though this distribution is less biased against extended clusters, it is still strongly peaked, which is further evidence for a characteristic size of YMCs in M83.

The distribution in Figure~\ref{rcore-hist} is somewhat different than that found by \cite{larsen2004}. In that study, the distribution of FWHM (equivalent twice our core radius) peaks at smaller values and declines more gradually to larger values than our distribution. We do, however, find a similar overall range of core radii. The core radii of 23 clusters in M31 span about the same range of values as Figure~\ref{rcore-hist} as well \citep{barmby2009}. Clusters in the Magellanic Clouds, however, seem to have larger average core radii than those found in nearby spirals \citep{elson1987,elson1991,mackey2003b,mackey2003a,glatt2009}.

A significant increase in the spread of core radii with increasing cluster age was observed for clusters in the LMC by \cite{elson1991} and \cite{mackey2003a}, in particular. \cite{sanroman2012} find a similar result, though they measure smaller core radii on average, in M33. In Figure~\ref{rcore-age}, we plot the core radii of our sample as a function of age (blue points), and overplot LMC clusters of similar age from \cite{mackey2003a} for comparison (gray diamonds). Clearly, the core radii of the LMC clusters in this age range are larger on average than the clusters in M83. We do not see a strong increase in the \textit{spread} of core radii in M83, but we do see that the core radius of the largest cluster increases with age, similar to the LMC. We find evidence for an overall trend of increasing core radius with increasing age. This trend has a slope of 0.24 $\pm$ 0.05, similar to the size-age trend in Figure~\ref{reff-age}. We do not see a bifurcation in Figure~\ref{rcore-age}, which \cite{mackey2003a} suggest is present in their data.

\subsection{EFF Index}
\label{effindex}

\begin{figure}[t]
\centering
\includegraphics[scale=0.6]{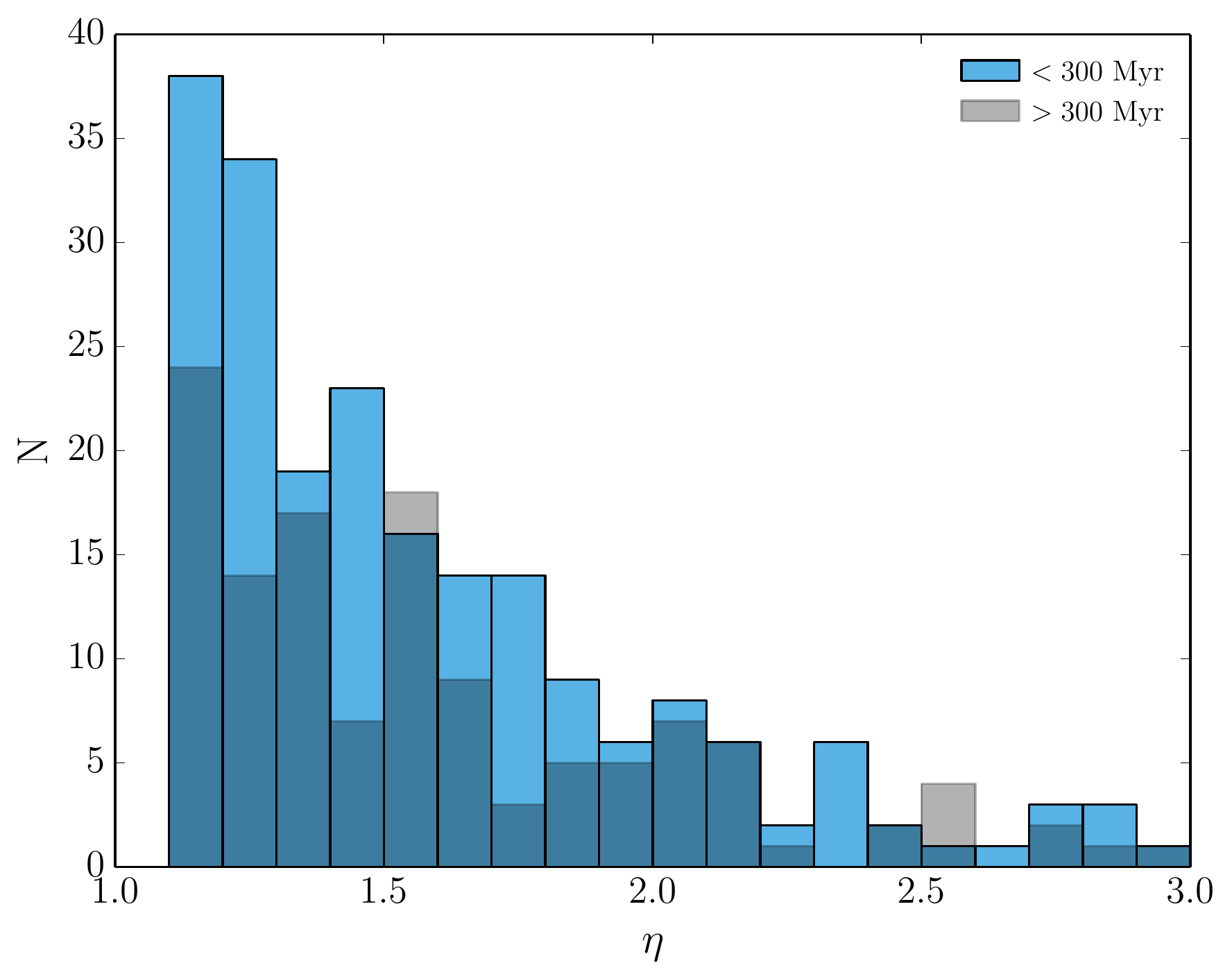}
\caption{Distribution of EFF power-law indices, $\eta$. Clusters younger than 300~Myr are plotted in blue and clusters older than 300~Myr are plotted in gray. Only clusters best-fit by $\eta\geq1.1$ are included. \label{index-hist}}
\end{figure}

We now turn to the EFF profile power-law index, $\eta$. Again, we have imposed an $\eta$ limit of $\eta\geq1.1$ for the reasons described in Section~\ref{rcore}. In Figure~\ref{index-hist}, we plot the distribution of best-fit $\eta$ values for our cluster sample. The distribution peaks at the lowest $\eta$ values, followed by a rapid falloff to higher $\eta$ values. The median of the young distribution is 1.5, similar to that found by previous studies \citep{elson1987,mackey2003a,larsen2004,mclaughlin2005,glatt2009}. A few objects are best-fit by $\eta>3.0$, up to $\eta\approx8.7$, but we choose to focus on the region between 1.0 and 3.0 where most of our clusters lie. This range is consistent with the previously reported best-fit $\eta$ values for young clusters \citep{elson1987, mackey2003a, larsen2004, mclaughlin2005, glatt2009}. Therefore, our choice to only fit relatively isolated clusters does not seem to bias our determination of the distribution of EFF power-law slopes. 

\begin{figure}[h]
\centering
\includegraphics[scale=0.6]{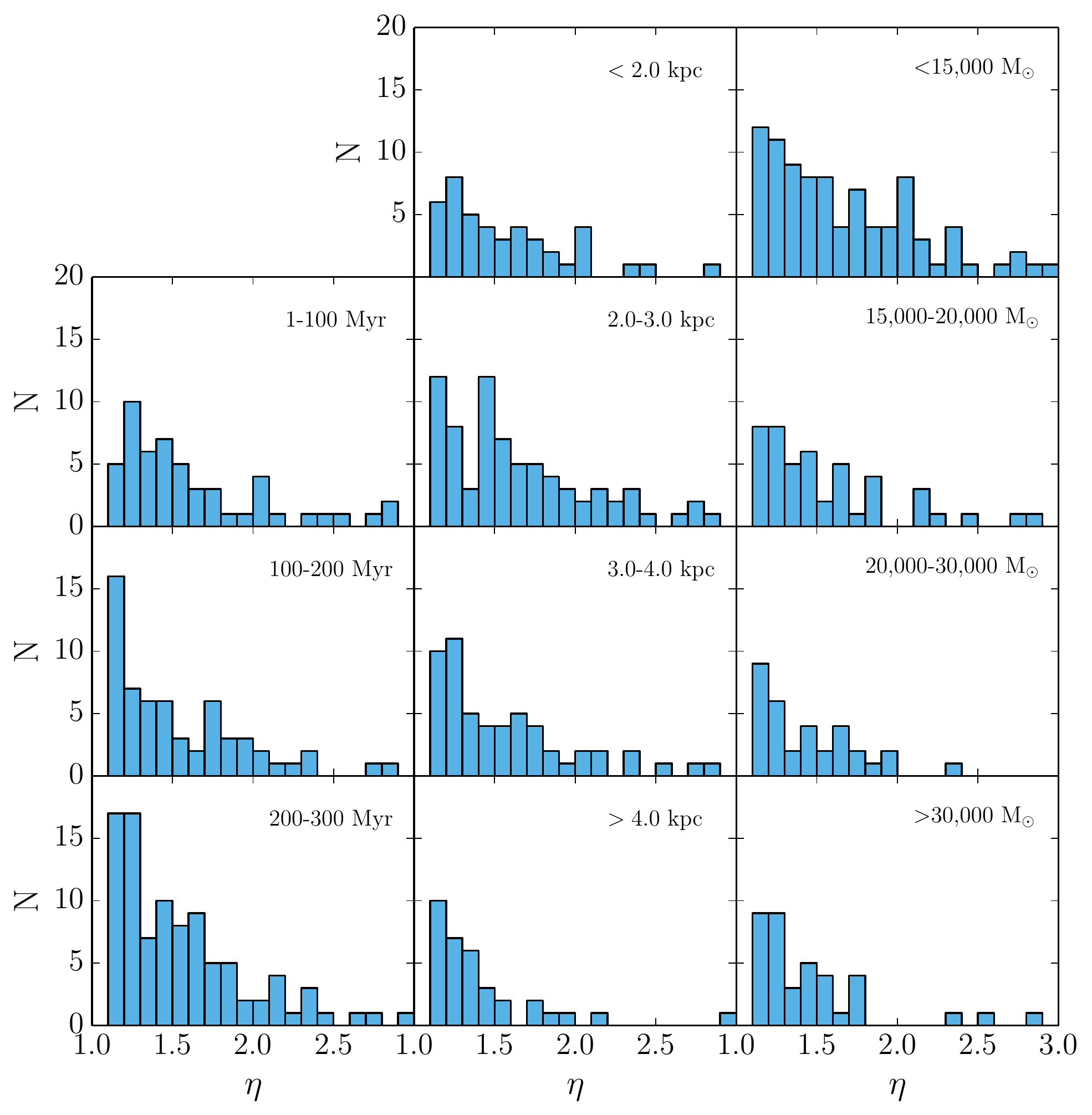}
\caption{Distribution of EFF power-law indices divided into three bins in age (left column), four bins in galactocentric distance (middle column), and four bins in mass (right column). Only clusters $\leq$300~Myr in age and best-fit by $\eta\geq1.1$ are included. \label{index-hist-grid}}
\end{figure}

We also look for any change in $\eta$ with other cluster properties. In Figure~\ref{index-hist-grid}, we divide the $\eta$ distribution into three bins in cluster age in the left column, four bins in galactocentric distance in the middle column, and four bins in mass in the right column. We perform Anderson-Darling tests to determine if the distributions within each column are significantly different from one another. This takes the form of rejecting or accepting the null hypothesis, i.e., the distributions in each column could be drawn from the same parent distribution. For instance, for the cluster age column, we perform Anderson-Darling tests comparing the first bin to the second, the second to the third, the first to the third, and all three together. The same is done for the other two columns, though the number of permutations increases when a fourth bin is included in the tests. We report the significance level with which the null hypothesis can be rejected for a given test. The $p$-values of each test that rejects the null hypothesis are less than or equal to the significance level. The Anderson-Darling test is ideal for comparing the distributions in Figure~\ref{index-hist-grid} because it is a non-parametric test, which means it is not necessary to know the shape of the true underlying distribution.

For the bins in the age column, we cannot reject the null hypothesis for any bin as tested against any other bin. Therefore, the $\eta$ distributions in each bin are consistent with being drawn from the same parent population. We find no evolution of $\eta$ with age.

For the bins in the galactocentric distance column, we can reject the null hypothesis at a significance level of 2\% for each permutation that tests the distribution in the fourth bin ($>$4.0~kpc) against those in the three other bins. Therefore, the probability of obtaining as extreme of a difference as we find between the fourth bin and the other three bins is 2\%. The top three bins are consistent with being drawn from the same parent population. We speculate that the reason for the distinctness of the distribution in the fourth bin is that it is sharply peaked at low values of $\eta$, and contains very few clusters best-described by $\eta>1.5$, unlike the other three bins. It is also possible that this difference is driven by the lack of clusters detected beyond 4~kpc.

For the bins in the mass column, we can reject the null hypothesis at a significance level of 4\% for tests that compare the first bin ($<$15,000~M$_{\odot}$) to the third and fourth bins. Every other comparison between bins results in no rejection of the null hypothesis, i.e., the other combinations are consistent with being drawn from the same parent distribution. There are many fewer clusters above 20,000~M$_{\odot}$ in our sample than clusters $<$15,000~M$_{\odot}$, so this difference in sample size could be driving the test results. Therefore, while the Anderson-Darling tests suggest there may be significant differences between some of the distributions in Figure~\ref{index-hist-grid}, the differences are rather small, and therefore should be further tested with studies of cluster populations in other galaxies.

\cite{larsen2004} find a tentative increase of the mean $\eta$ (which they call $\alpha$) with cluster age, although this is mostly due to their youngest age bin containing a larger fraction of clusters with $\eta<1.0$ than the other age bins. Since we cannot determine which of our clusters are truly best described by $\eta<1.0$, we cannot test this result. Using the data presented in \cite{mackey2003a} and \cite{mclaughlin2005} for LMC and SMC clusters, we looked for a relationship between age and $\eta$ and found none, although no clusters in those samples are consistent with $\eta <1.0$.

\section{Discussion}
\label{discussion}

\subsection{Drivers of Cluster Structural Evolution}
\label{drivers}

The expulsion of leftover gas from formation appears to be completed within the first 4~Myr of a star cluster's lifetime \citep{hollyhead2015}. Excluding the two youngest ($\approx$5~Myr) objects, the clusters in our sample are therefore likely to be gas-free. The radii we measure for these clusters cannot be considered their initial radii, since clusters are expected to expand significantly between formation and the end of gas expulsion \citep[e.g.,][]{portegieszwart2010}. Similarly, the initial light profile slopes may have changed during this early phase of dynamical evolution. While we cannot know the initial configurations of our clusters, we can characterize their structural evolution between $\sim$10 to 300~Myr.

The dominant \textit{internal} mechanism affecting the structure of clusters 10 to 300~Myr in age is mass loss due to stellar evolution \citep[e.g.,][]{portegieszwart2010}. If mass is lost slowly as compared to the dynamical time, an isolated cluster can maintain virial equilibrium while it gradually expands. Generally, mass loss can be assumed to be slow for massive clusters because even if a single star loses the majority of its mass very quickly, the amount of mass lost by the cluster is a relatively small fraction of the total \citep{heggiehut2003}. If the cluster is initially mass-segregated, this expansion will occur mostly in the core, since the massive stars in the core will lose relatively more mass than the rest of the population. On the other hand, the effective radius of a tidally-limited cluster should contract as it loses mass, but its core will expand \citep{heggiehut2003}. Therefore, we would expect to see an expansion of core radii, and an expansion or contraction of effective radius with age if mass loss through stellar evolution dominates over external forces. Since we find an expansion of the mean effective and core radii with cluster age in Figures~\ref{reff-age} and \ref{rcore-age}, it seems likely that our clusters are not tidally-limited.

Two-body relaxation also changes cluster structure, but typically becomes dominant over longer timescales than we encounter here \citep{portegieszwart2010}. We test this by calculating approximate half-mass relaxation times of the clusters using Equation~14.13 from \cite{heggiehut2003}. For a 100~Myr old, 10$^4$~M$_{\odot}$ star cluster with an effective radius of 3~pc, the present-day, half-mass relaxation time is $\approx$200~Myr. Here we assume that the three-dimensional half-mass radius is equal to $4/3\times$ the effective radius \citep{spitzer1987}. We also assume that the average stellar mass is 0.65~M$_{\odot}$, which was estimated using a Kroupa IMF over a mass range of 0.1 to an approximate turn-off mass for a 100~Myr cluster, 6~M$_{\odot}$. The relaxation time is strongly dependent on the cluster radius such that more compact clusters have shorter relaxation times. Given the distributions of radii and masses in our sample, the relaxation times range from a few tens of Myr to a few Gyr, with a peak around 300~Myr. About 80\% of our clusters are younger than one relaxation time, and the rest are about equal to or a few times older than their relaxation times. However, given that clusters are expected to expand drastically shortly after their formation, and that we observe a modest expansion during the first few hundred Myr of cluster evolution, it is likely that the half-mass relaxation times were shorter in the past. Whether two-body relaxation was a dominant process immediately after their formation, we cannot say, but it appears that the majority of the clusters in our sample are not strongly affected by two-body relaxation at the present day.

Two \textit{external} processes that may influence cluster structure are the tidal field of M83 and interactions with giant molecular clouds (GMCs). As mentioned previously, clusters that are tidally-limited behave differently during the period of significant stellar mass loss than those that are not. A cluster that is not tidally-limited can evolve essentially as an isolated system. The lack of a strong relationship between mean effective radius and galactocentric distance in Figures~\ref{reff-dgc} and \ref{reff-dgc-agebins} suggests our sample is not strongly affected by tides. If it was, we would expect the clusters to become much larger with increasing galactocentric distance as the tidal field becomes weaker \citep[e.g.,][]{madrid2012}. One potential signature of tidal influence on our sample is the tentative preference for shallower light profiles further from the galaxy center in Section~\ref{effindex}.

A further test of the influence of the tidal field on the clusters is to calculate approximate Jacobi radius, which defines the zero-velocity surface of a cluster in a tidal field. Stars within the Jacobi radius can be said to ``belong'' to the cluster. The Jacobi radius is defined by Equation~9 in \cite{portegieszwart2010},
\begin{equation}
\label{jacobi}
r_{\mathrm{J}} = \left(\frac{GM}{2\omega^2}\right)^{1/3},
\end{equation}
where $M$ is the mass of the cluster and $\omega$ is the angular velocity of the cluster around the galactic center. We estimate $\omega$ by assuming the angular velocity of molecular gas in the disk of M83 matches, to first-order, that of star clusters located at the same galactocentric distance. We found an approximate linear relation between $\omega$ of CO in M83's disk and galactocentric distance from the top panel of Figure~4 in \cite{lundgren2004}. Using this relation, we estimate the Jacobi radius of each cluster in our sample, and find that they range in size from $\sim$14 to $\sim$43~pc. Above $r_{\mathrm{hm}}/r_{\mathrm{J}}\approx0.15$ \citep{henon1961} or $\approx$0.2 \citep{alexander2014}, a cluster fills its Roche volume. We find Jacobi to half-mass radii ratios between $\approx$0.02 to $\approx$0.5. About 68\% of our clusters have $r_{\mathrm{hm}}/r_{\mathrm{J}}\leq0.2$, and 43\% have $r_{\mathrm{hm}}/r_{\mathrm{J}}\leq0.15$. Therefore, based on rough estimates of the Jacobi radii, our sample is straddling the line between being influenced by tides and not being influenced by tides, i.e., some of our clusters are likely underfilling their Roche volumes while others are filling them. Comparing these results to cluster age, we find that larger Jacobi to half-mass radii ratios are more common in older clusters, so perhaps clusters become tidally-limited as they age.

\cite{gieles2006c} studied the effect of interactions between GMCs and star clusters on the mass loss rate and disruption time of clusters. They show that a 10$^4$~M$_{\odot}$ cluster in the solar neighborhood would lose enough mass to be disrupted in about 2~Gyr. Each interaction leads to an energy gain by the cluster, though the magnitude of that gain is dependent on the characteristics of the cloud, the cluster, and the interaction itself. We would expect an energy gain by the cluster to cause expansion (due to a shallower potential well), but it is unclear if the degree of tidal filling would have an impact on this expansion. Assuming expansion occurs, and that it takes place at a similar rate to mass loss, the clusters in our sample may be significantly affected by GMC interactions over the age range we study.

\subsection{Similarity of Cluster Sizes}
\label{sim}

In Section~\ref{dist_reff}, we compare the radius distribution of our cluster sample with that of other samples in the literature. Our sample is one of the most extensive in a single galaxy for which the slope of the light profile ($\eta$) was left as a free parameter, and yet we find essentially the same distribution as found in many other studies. As many of these studies note, star clusters of a wide range in age, mass, and environment all seem to be $\approx$3~pc in size. Even GCs, which have undergone a Hubble time of dynamical evolution and been affected by different physical processes (e.g., two-body relaxation, tidal shocks, core collapse), are similar in size to YMCs.

In Section~\ref{dist_reff}, we note the existence of populations of extended clusters (both young and old) in the Magellanic Clouds, M31, M101, several S0 galaxies, and even the Milky Way, calling into question the conclusion that cluster sizes are relatively constant. In the Magellanic Clouds, clusters that have been carefully measured are, on average, larger than the clusters in M83 (see Figure~\ref{rcore-age}). However, this well-defined sample contains only 63 objects from a combined population of a few thousand in both Clouds \citep{glatt2010}. The larger average size of Magellanic Cloud clusters could therefore be a selection effect, i.e., more compact clusters simply have not been studied at the same level of detail. Another possibility is that local environment somehow does play a role in determining cluster structure. Studies of larger samples of Magellanic Cloud clusters and of clusters in other dwarf galaxies are needed to answer this question.

A number of extended clusters do exist in M83, as shown in the bottom panel of Figure~\ref{reff-hist}, although we cannot precisely determine their sizes. We note, however, that the number of extended clusters is relatively small as compared to the rest of the population. This seems to be the case in most studies that find extended clusters in other galaxies; clusters $\approx$3~pc in size outnumber those of much larger size \citep[e.g.,][]{larsen2000, simanton2015}. These extended clusters appear to represent a tail in the size distribution rather than a separate population, and the majority of star clusters seem to have a characteristic compact size of $\approx$3~pc. Faint fuzzies may be an exception, or could simply be old open clusters that are easier to detect in S0 galaxies than star-forming spirals \citep{chiessantos2013}. Of course, deeper surveys are needed to determine if large numbers of faint, extended clusters in nearby spiral galaxies have gone undetected.

Perhaps the most surprising feature of the similarity between cluster sizes is the weak mass-radius relationship, although we find that this relationship strengthens as clusters age (Section~\ref{size-mass}). GMCs have a fairly strong mass-radius relationship, on the order of $r_{\mathrm{GMC}}\propto M_{\mathrm{GMC}}^{0.5}$ \citep{larson1981}. It has been argued extensively that some process, either during cluster formation or subsequent evolution, must flatten the GMC mass-radius relationship. \cite{ashman2001} suggest that if the star formation efficiency of GMCs depends on their binding energy, then the clusters that form from the GMCs will have no mass-radius relationship. This, however, implies that the subsequent dynamical evolution of the clusters cannot have a strong effect on their sizes. \cite{gieles2010b} explore the possibility that all self-gravitating stellar systems have the same mass-radius relation upon formation and lower mass clusters have moved away from this relation due to expansion driven by stellar mass loss and hard binaries. This idea is supported by studies briefly discussed in Section~\ref{size-mass}, which find a clear mass-radius correlation for objects $\gtrsim$10$^6$~M$_{\odot}$. In addition, the theory presented in \cite{gieles2010b} implies a similar rate of expansion as we observe in M83. However, the mass-radius relation we find becomes stronger with increasing cluster age, which is inconsistent with the suggestion by \cite{gieles2010b} that the initial mass-radius relation should degrade over time as low-mass clusters expand.

\section{Conclusions}
\label{conclusions}

We have measured the effective radii, core radii, and EFF power-law index of a large sample of YMCs in M83. To do this, we used F547M/F555W images from seven HST pointings that cover the inner $\sim$6~kpc of M83. Our cluster sample was selected from the catalogue presented by \cite{silvavilla2014}, and consists of relatively isolated, well-resolved, massive ($\geq$10$^4$~M$_{\odot}$) clusters. We use GALFIT to fit the two-dimensional light profiles of our clusters assuming an EFF power-law profile. The main results from this study are summarized as follows:

\begin{itemize}

\item[1.] The distribution of effective radii is well-represented by a lognormal distribution with a median at $\approx$2.5~pc and a range of $\approx$0.3 to 10~pc (Figure~\ref{reff-hist}). Our sample is likely incomplete in both very compact and very extended clusters, but the $\approx$3~pc peak is present in our data even when including more extended objects. We suggest this is strong evidence for a characteristic size of YMCs. Our effective radius distribution is very similar to that found by many studies in the literature of both YMCs and GCs in the Milky Way and other nearby galaxies.

\item[2.] We find that the mean effective radii, calculated in bins of 0.3 dex in age, tend to increase with cluster age (Figure~\ref{reff-age}). We interpret this as evidence of expansion over the first few hundred Myrs of evolution. This trend has a power-law slope of 0.26 $\pm$ 0.07. Other studies of young massive clusters in nearby galaxies typically find similar shallow relationships between age and size, notably in M83 \citep{bastian2012a}, or no obvious trend at all. 

\item[3.] The effective radii of clusters in our sample do not show a strong correlation with distance from the center of M83. A weak trend is visible in Figure~\ref{reff-dgc}. When we separate our clusters into three age bins to remove the age-size correlation in Figure~\ref{reff-dgc-agebins}, we find little to no correlation between mean effective radius and galactocentric distance, except perhaps for the youngest clusters. Several other studies of young massive clusters have found similar results. In studies of GCs, however, some systems display a strong correlation between size and distance, such as the Milky Way, and others do not.

\item[4.] We find a slight increase in effective radius with cluster mass, with a power-law slope of 0.3 $\pm$ 0.1 (Figure~\ref{reff-mass}). If we separate the clusters into bins of cluster age, we find the mass-radius relation becomes stronger with age (Figure~\ref{reff-mass-agebin}). Many studies in the literature have found little to no correlation between mass and radius for both YMCs and GCs.

\item[5.] The distribution of core radii is fairly lognormal in shape with a median at $\approx$1.3~pc and a range of $\approx$0.1 to 4~pc (Figure~\ref{rcore-hist}). This distribution is less biased against extended clusters because of the relaxed $\eta$ limit, and it is still strongly peaked. In addition, this distribution is somewhat similar to those measured for clusters in nearby galaxies, except for perhaps the Magellanic Clouds. We also find an overall trend of increasing core radius with increasing age (Figure~\ref{rcore-age}), which is very similar to the effective radius trend with age. This trend has a power-law slope of 0.24 $\pm$ 0.05.

\item[6.] We measure the power-law slope of the EFF light profile, $\eta$, for each cluster. The distribution of $\eta$ values peaks in the lowest bin and rapidly falls off to higher values (Figure~\ref{index-hist}). The median $\eta$ is 1.5, and the majority of the clusters are best described by $\eta<3.0$. Other studies find a similar range of $\eta$ values as we do here. In an attempt to find change in $\eta$ with other cluster parameters, we divide the $\eta$ distribution into bins of cluster age, galactocentric distance, and mass (Figure~\ref{index-hist-grid}). While statistical tests suggest significant differences exist between some of the bins in galactocentric distance and mass, the differences are not obvious. Evolution of $\eta$ should be further tested with cluster population studies in more galaxies.

\item[7.] The expansion of our clusters with age is most likely due to mass loss by stellar evolution or GMC interactions. In addition, the tidal field of M83 appears to play a minor role in setting the size of young clusters in our sample. If our clusters were strongly limited by tides, we would expect to see a contraction of the effective radius with age instead of an expansion. We would also expect to see a stronger mass-radius relation, of order $r_{\mathrm{eff}}\propto M^{1/3}$ (Equation~\ref{jacobi}). We do find a slope of $\approx$0.5 in the 200-300~Myr bin in Figure~\ref{reff-mass-agebin}, which may be evidence of clusters becoming tidally-limited as they age. This is supported by a slight increase of the average $r_{\mathrm{hm}}/r_{\mathrm{J}}$ with cluster age. However, in conjunction with the lack of a strong relationship between radius and galactocentric distance in Figure~\ref{reff-dgc-agebins}, the evidence suggests that the structure of younger clusters with masses $\geq$10$^4$~M$_{\odot}$ does not appear to be strongly affected by galactic tides. 

\end{itemize}

\section{Acknowledgements}

J.E. Ryon gratefully acknowledges the support of the National Space Grant College and Fellowship Program and the Wisconsin Space Grant Consortium. This research has made use of the NASA/IPAC Extragalactic Database (NED) which is operated by the Jet Propulsion Laboratory, California Institute of Technology, under contract with the National Aeronautics and Space Administration. Based on observations made with the NASA/ESA Hubble Space Telescope, obtained from the data archive at the Space Telescope Science Institute. STScI is operated by the Association of Universities for Research in Astronomy, Inc. under NASA contract NAS 5-26555. J.E. Ryon thanks the staff of the Australian Astronomical Observatory for her visit, during which a part of this work was undertaken. I.S. Konstantopoulos is the recipient of a John Stocker Postdoctoral Fellowship from the Science and Industry Endowment Fund (Australia). This research is supported by the Science and Industry Endowment Fund (Australia). E. Silva-Villa acknowledges Estrategia de Sostenibilidad 2014-2015 de la Universidad de Antioquia.

\bibliography{m83}

\end{document}